\documentclass[aps,pra,reprint,twocolumn,showpacs]{revtex4-1}

\usepackage{graphicx}

\usepackage{amsmath}

\usepackage{amssymb}

\usepackage{natbib}
\usepackage{bibentry}

\newcommand{\mbf}[1]{\mathbf{#1}}

\newcommand{\rbr}[1]{\ensuremath{\left(#1\right)}}
\newcommand{\sbr}[1]{\ensuremath{\left[#1\right]}}
\newcommand{\an}[1]{\ensuremath{\left\langle#1\right\rangle}}
\newcommand{\set}[1]{\left\{#1\right\}}

\newcommand{\br}{\ensuremath{\mbf{r}}}
\newcommand{\ofr}{\ensuremath{\rbr{\mbf{r}}}}
\newcommand{\dr}{\ensuremath{d\mbf{r}}}

\newcommand{\abs}[1]{\left|#1\right|}

\usepackage[dvipsnames]{color}

\begin{document}


\title{Metric space approach to potentials and its relevance to density functional theory}


\author{P. M. Sharp}
\email{pms510@york.ac.uk}
\affiliation{Department of Physics and York Centre for Quantum Technologies, University of York, York, YO10 5DD, United Kingdom}
\altaffiliation{Present address: Department of Chemistry, University of Liverpool, Crown Street, Liverpool, L69 7ZD, United Kingdom}
\author{I. D'Amico}
\email{irene.damico@york.ac.uk}
\affiliation{Department of Physics and York Centre for Quantum Technologies, University of York, York, YO10 5DD, United Kingdom\\
and Instituto de Fısica de Sao Carlos, Universidade de Sao Paulo, Caixa Postal 369, 13560-970 Sao Carlos, SP, Brazil}

\date{\today}

\begin{abstract}
External potentials play a crucial role in modelling quantum systems, since, for a given inter-particle interaction, they define the system Hamiltonian.
We use the metric space approach to quantum mechanics to derive, from the energy conservation law, two natural metrics
for potentials. We show that these metrics are well defined for physical potentials, regardless of whether the system is in an eigenstate or if the potential is bounded.
In addition, we discuss the gauge freedom of potentials and how to ensure that the metrics preserve physical relevance.
Our metrics for potentials, together with the metrics for wavefunctions and densities from \citetext{I. D'Amico, J. P. Coe, V. V. Fran\c{c}a, and K. Capelle,
Phys. Rev. Lett. 106, 050401 (2011)} paves the way for a comprehensive study of the two fundamental theorems of Density Functional Theory.
We explore these by analysing two many-body systems for which the related exact Kohn-Sham systems can be derived. First we consider the information provided by each of the metrics,
and we find that the density metric performs best in distinguishing two many-body systems. Next we study for the systems at hand the one-to-one relationships among potentials,
ground state wavefunctions, and ground state densities defined by the Hohenberg-Kohn theorem as relationships in metric spaces.
We find that, in metric space, these relationships are monotonic and incorporate regions of linearity, at least for the systems considered.
Finally, we use the metrics for wavefunctions and potentials in order to assess quantitatively how close the many-body and Kohn-Sham systems are:
We show that, at least for the systems analysed, both metrics provide a consistent picture, and for large regions of the parameter space
the error in approximating the many-body wavefunction with the Kohn-Sham wavefunction lies under a threshold of 10\%.
\end{abstract}


\maketitle


\section{Introduction}

Density functional theory (DFT) is one of the most widely used methods for performing quantum mechanical analysis of many-body systems.
DFT is founded upon two core theorems. The first of these is the Hohenberg-Kohn theorem~\citep{HK1964}, which demonstrates, for ground states,
that the many-body wavefunction, the external potential, and the density are uniquely determined by each other:
\begin{equation} \label{HK_map}
 V\rbr{\mbf{r},\mbf{r}_{2},\ldots,\mbf{r}_{N}} \rightleftharpoons \psi\rbr{\mbf{r},\mbf{r}_{2},\ldots,\mbf{r}_{N}} \rightleftharpoons \rho\ofr.
\end{equation}
Therefore, wavefunctions, potentials, and expectation values of any operator can, in principle, be written as functionals of the ground-state density.
The Hohenberg-Kohn theorem applies for any given strength of the interaction between the particles. Thus, in the second core theorem of DFT, Kohn and Sham
recognised that the many-body system of interacting particles can be described by an auxiliary system of \textit{non-interacting} particles, in a different external
potential (the Kohn-Sham potential), that produces the same ground-state density~\citep{KS1965}. Since the Kohn-Sham particles are non-interacting, the wavefunction for this system is composed of
single-particle orbitals, found by solving a system of single-particle equations, the Kohn-Sham equations.
The solution of these equations thus provides a method to obtain the many-body ground state density that bypasses the many-body wavefunction (the Kohn-Sham scheme)~\citep{KS1965}.

These two theorems are sufficient to construct DFT in a formal way; however, there are open questions with regards to both of them.
Although the Hohenberg-Kohn theorem guarantees a one-to-one relationship between
potentials and ground-state wavefunctions, as well as ground-state wavefunctions and ground-state densities, it offers no prescription on how these
wavefunctions or potentials are produced given a particular density. For the Kohn-Sham scheme, although it is known that the Kohn-Sham potential is constructed
from the sum of external, Hartree, and exchange-correlation potentials, the exchange-correlation component is generally unknown
and hence must be approximated when DFT calculations are implemented practically. There are numerous approximations to the exchange-correlation potential that cover a
wide range of sophistication and complexity~\citep{Burke2012}, and the suitability of an approximation usually depends on the problem studied.

In this work, we apply the metric space approach to quantum mechanics~\citep{D'Amico2011,Sharp2014,Sharp2015} to potentials in order to gain insight into the
two fundamental theorems of DFT. First, we use the general procedure from Ref.~\citep{Sharp2014} to derive two metrics for external potentials.
These metrics will complement the metrics for wavefunctions and densities derived in Ref.~\citep{D'Amico2011} and ensure that we have metrics for each of the fundamental
physical quantities associated to DFT. We will then revisit the Hohenberg-Kohn theorem. This was first studied with the metric space approach to quantum mechanics in
Ref.~\citep{D'Amico2011}, where only the second part of Eq.~(\ref{HK_map}), concerning ground-state wavefunctions and densities, was studied.
Now, with the external potential metrics, we will extend the study to incorporate the first part of Eq.~(\ref{HK_map}), which establishes a unique map between the
external potential and the ground-state wavefunction. We will then turn our attention to the Kohn-Sham scheme. By studying model systems for which the Kohn-Sham quantities
can be determined exactly, we will use our metrics to quantify the differences between many-body and Kohn-Sham quantities. We will use atomic units $\rbr{\hbar=m_{e}=e=1/4\pi\epsilon_{0}=1}$ throughout this paper.

\section{Deriving Metrics for Potentials}\label{sec:metric}

In order to derive a metric for external potentials, we use the metric space approach to quantum mechanics~\citep{D'Amico2011,Sharp2014,Sharp2015}, which allows us to derive metrics
from conservation laws of the form
\begin{equation} \label{conservation}
 \int\abs{f\rbr{\mbf{x}}}^{p} d\mbf{x} = c,
\end{equation}
where $c$ is a finite, positive constant. Equation~(\ref{conservation}) has the form of an $L^p$ norm, from which a metric can be derived in a standard way.
As these metrics then naturally descend from the physical conservation laws, we refer to them as ``natural'' metrics for the related physical functions.
A metric is a function that assigns a distance between two elements of a set and is subject to the axioms~\citep{Sutherland2009,Megginson1998}
\begin{align}
D\rbr{x,y} &\geqslant 0\ \text{and}\ D\rbr{x,y}=0 \iff x=y, \label{axiom1}\\
D\rbr{x,y} &= D\rbr{y,x}, \label{axiom2}\\
D\rbr{x,y} &\leqslant D\rbr{x,z}+D\rbr{z,y}, \label{axiom3}
\end{align}
for all elements $x,y,z$ in the set. A set with an appropriate metric defined on it is called a metric space.

In time-independent quantum mechanics, the system energy is conserved and it is given by the expectation value
\begin{equation} \label{energy_cons0}
 \int\ldots\int\psi^{*}\rbr{\br_{1},\ldots,\br_{N}}\hat{H}\psi\rbr{\br_{1},\ldots,\br_{N}} d\mbf{r}_{1}\ldots d\mbf{r}_{N} = EN,
\end{equation}
where
\begin{equation} \label{hamiltonian}
\hat{H}= -\sum_{i=1}^{N}\frac{1}{2}\nabla_{i}^{2} + \sum_{j<i}^{N} U\rbr{\br_{i},\br_{j}} + \sum_{i=1}^{N} v\rbr{\mbf{r}_{i}},
\end{equation}
is the system Hamiltonian, where $V=\sum_{i=1}^{N} v\rbr{\mbf{r}_{i}}$ is the external potential and $\psi\rbr{\br_{1},\ldots,\br_{N}}$ is the system state.
We have followed Ref.~\citep{D'Amico2011} and normalised the many-body wavefunction $\psi\rbr{\br_{1},\ldots,\br_{N}}$ to the particle number $N$.
In the following we will concentrate on the Coulomb particle-particle interaction $U\rbr{\br_{i},\br_{j}}=1/|\br_{i}-\br_{j}|$, though the results
are valid for a general form of $U\rbr{\br_{i},\br_{j}}$. In Eq.~(\ref{hamiltonian}) and the following analysis we focus on electronic systems, as is often
done in studies involving DFT when invoking the Born-Oppenheimer approximation. However, our results can be extended to include nuclear terms in the Hamiltonian,
which we demonstrate in the Appendix. The derivations in the Appendix can be straightforwardly extended to more complex systems comprising various particles and/or species,
such as systems including electrons and different ionic species.

We will now derive metrics for the external potential from Eq.~(\ref{energy_cons0}) by applying the metric space approach to quantum mechanics.
We start by performing some simple algebra and rewrite Eq.~(\ref{energy_cons0}) in the following two forms:
\begin{align} \label{energy_cons1}
\int&\ldots\int \sum_{i=1}^{N}{\sbr{-\frac{1}{2}\psi^{*}\nabla_{i}^{2}\psi+\sum_{j<i}^{N}{\frac{\abs{\psi}^{2}}{\abs{\mbf{r}_{i}-\mbf{r}_{j}}}}+\abs{\psi}^{2}v\rbr{\mbf{r}_{i}}}}\nonumber\\
&\times \dr_{1}\ldots d\mbf{r}_{N} = EN
\end{align}
and
\begin{equation}\label{energy_cons2}
\int N\sbr{\tau\rbr{\mbf{r}}+ \frac{1}{2}\int \dr_{1}\frac{g\rbr{\mbf{r},\mbf{r}_{1}}}{\abs{\mbf{r}-\mbf{r}_{1}}}+v\rbr{\mbf{r}}\rho\rbr{\mbf{r}}}\dr=EN.
\end{equation}
Here, we have used the definitions
\begin{equation}
\tau\rbr{\mbf{r}}\equiv \frac{1}{2}\int\ldots\int\abs{\nabla_{\mbf{r}}\psi\rbr{\mbf{r},\mbf{r}_{2},\ldots,\mbf{r}_{N}}}^{2}\dr_{2}\ldots d\mbf{r}_{N}\geqslant0\label{kinetic_density}\\
\end{equation}
for the kinetic energy density,
\begin{equation}
g\rbr{\mbf{r}_{1},\mbf{r}_{2}}\equiv\rbr{N-1}\int\ldots\int\abs{\psi\rbr{\mbf{r}_{1},\mbf{r}_{2},\ldots,\mbf{r}_{N}}}^{2}\dr_{3}\ldots d\mbf{r}_{N}\geqslant0\label{2_part_corr}\\
\end{equation}
for the two-particle correlation function, and
\begin{equation}
\rho\rbr{\mbf{r}}\equiv \int\ldots\int\abs{\psi\rbr{\mbf{r},\mbf{r}_{2},\ldots,\mbf{r}_{N}}}^{2}\dr_{2}\ldots d\mbf{r}_{N}\geqslant0,\label{1_part_density}\\
\end{equation}
for the single-particle density. To derive Eq.~(\ref{kinetic_density}), we have used that for any $i=1\ldots N$
\begin{align} \label{ke_relation}
 -\frac{1}{2}\int\psi^{*}\nabla_{i}^{2}\psi\dr_{i}&=-\frac{1}{2}\sbr{\psi^{*}\nabla_{i}\psi}_{\br_{i}\rightarrow\infty}+\frac{1}{2}\int\sbr{\rbr{\nabla_{i}\psi^{*}}\cdot\rbr{\nabla_{i}\psi}}\dr_{i}\nonumber\\
 &=\frac{1}{2}\int\abs{\nabla_{i}\psi}^{2} \dr_{i},
\end{align}
as $\psi\rightarrow 0$ when $\br_{i}\rightarrow\infty$. This also shows that the kinetic term in Eq.~(\ref{energy_cons1}) is positive.

To derive ``natural'' metrics, we must ensure that the conservation laws Eqs.~(\ref{energy_cons1}) and~(\ref{energy_cons2}) can be written in the form of Eq.~(\ref{conservation}), so,
after taking the absolute value of their left and right sides, we need to demonstrate that the integrands in their left-hand sides always have the same sign throughout the corresponding domains.
From previous considerations, the parts of these integrands corresponding to the kinetic and particle-particle interaction terms, for both Eqs.~(\ref{energy_cons1}) and~(\ref{energy_cons2}),
are positive semi-definite everywhere, so we need only to consider the external potential term.

Although we cannot guarantee the sign of $v\rbr{\br}$, we can make use of a gauge transformation. If the potential is modified by a constant, $v\rbr{\br} \rightarrow v\rbr{\br}+c$,
then the solution to the Schr\"{o}dinger equation is unaffected. Thus, for potentials with a lower bound, we can choose a constant $c$ such that the potential term (and hence the overall
integrand) in Eqs.~(\ref{energy_cons1}) and~(\ref{energy_cons2}) is positive semi-definite everywhere
\footnote{We will consider the important case of a bare, attractive Coulomb potential in Sec.~\ref{sec:coulomb}}.

With this in mind we can rewrite Eqs.~(\ref{energy_cons1}) and (\ref{energy_cons2}) as
\begin{align} \label{pot_norm1}
\int&\ldots\int\abs{\sum_{i=1}^{N}{\sbr{\frac{1}{2}\abs{\nabla_{i}\psi}^2+\sum_{j<i}^{N}{\frac{\abs{\psi}^{2}}{\abs{\mbf{r}_{i}-\mbf{r}_{j}}}}+\abs{\psi}^{2}\sbr{v\rbr{\mbf{r}_{i}}+c}}}}\nonumber\\
&\times\dr_{1}\ldots d\mbf{r}_{N} = \abs{\rbr{E+c}N},
\end{align}
and
\begin{align} \label{pot_norm2}
\int&\abs{N\sbr{\tau\rbr{\mbf{r}}+\frac{1}{2}\int\dr_{1}\frac{g\rbr{\mbf{r},\mbf{r}_{1}}}{\abs{\mbf{r}-\mbf{r}_{1}}}+\sbr{v\rbr{\mbf{r}}+c}\rho\rbr{\mbf{r}}}}\dr\nonumber\\
&=\abs{\rbr{E+c}N}.
\end{align}
Given that both Eq.~(\ref{pot_norm1}) and Eq.~(\ref{pot_norm2}) are of the sought form~(\ref{conservation}), we can apply the metric space approach to quantum mechanics~\citep{Sharp2014}
and derive the corresponding metrics, which read
\begin{align}
&D_{v_{1}}=\int\ldots\int\abs{f_{1}-f_{2}} \dr_{1}\ldots d\mbf{r}_{N},\label{pot_metric1}\\
&D_{v_{2}}=\int\abs{h_{1}-h_{2}}\dr,\label{pot_metric2}
\end{align}
where
\begin{align}
f&\rbr{\mbf{r}_{1},\ldots,\mbf{r}_{N}}\nonumber\\
&\equiv\sum_{i=1}^{N}\set{\frac{1}{2}\abs{\nabla_{i}\psi}^2+\sum_{j<i}^{N}{\frac{\abs{\psi}^{2}}{\abs{\mbf{r}_{i}-\mbf{r}_{j}}}}+\abs{\psi}^{2}\sbr{v\rbr{\mbf{r}_{i}}+c}},
\end{align}
and
\begin{equation} \label{h_r}
h\rbr{\mbf{r}}\equiv N\sbr{\tau\rbr{\mbf{r}}+\frac{1}{2}\int\dr_{1}\frac{g\rbr{\mbf{r},\mbf{r}_{1}}}{\abs{\mbf{r}-\mbf{r}_{1}}}+\sbr{v\rbr{\mbf{r}}+c}\rho\rbr{\mbf{r}}}.
\end{equation}
$D_{v_{1}}$ and $D_{v_{2}}$ apply to both the case in which the system is in an eigenstate and when a more general system state is considered, as demonstrated below.

We note that both $\tau\rbr{\mbf{r}}$ and $g\rbr{\mbf{r},\mbf{r}_{1}}$ are uniquely defined by the many-body
wavefunction, $\psi\rbr{\br_{1},\ldots,\br_{N}}$. When the system is in an eigenstate, and for a given particle number and many-body interaction, the time-independent Schr\"{o}dinger equation
shows that the many-body wavefunction is uniquely determined by the external potential $v\rbr{\mbf{r}}$. Hence, every term in the integrands of both Eq.~(\ref{pot_norm1}) and Eq.~(\ref{pot_norm2})
(and hence in the related metrics) can be uniquely written as a functional of the external potential so that $f=f\sbr{v}$ and $h=h\sbr{v}$.
This demonstrates that Eqs.~(\ref{pot_norm1}) and~(\ref{pot_norm2}) indeed define two norms (and hence metrics) for the external potential $v\rbr{\mbf{r}}$.
It is simple to show that, when comparing the same two systems, $D_{v_{2}}<D_{v_{1}}$.

We note that the metric $D_{v_{2}}$ is well defined for comparing systems with different numbers of particles because it relies on a single-particle quantity, the function
$h\ofr$ defined in Eq.~(\ref{h_r}). The metric $D_{v_{1}}$ instead is well defined here only for systems with the same number of particles, $N_{1}=N_{2}$. The issue of defining
$D_{v_{1}}$ for systems with different numbers of particles is an open problem related to the fact that the wavefunction is a many-particle quantity. This issue has been discussed
previously with reference to $D_{\psi}$~\citep{Arthacho2011,D'Amico2011b}.

When considering a system with a \textit{time-independent} Hamiltonian but not in an eigenstate, conservation of energy applies to the time evolution of this state. In this case we can still
consider the norms (\ref{pot_norm1}) and (\ref{pot_norm2}) as derived from the conservation of energy. However, now the system state at any time $t$, $\psi\rbr{t}$, will still be
determined by the external potential $v\rbr{\mbf{r}}$, but together with the initial condition $\psi\rbr{t=0}$. The norms (\ref{pot_norm1}) and (\ref{pot_norm2}) will then still represent norms
for the external potential $v\rbr{\mbf{r}}$, and at any time $t$, but {\it given the initial state $\psi\rbr{t=0}$}. This condition mirrors the condition for uniqueness of the relationship
between the potential and the wavefunction $v\rbr{t}\longleftrightarrow\psi\rbr{t}$ as set in the core theorems of Time-Dependent DFT~\citep{Ullrich2013}, where indeed this uniqueness is subject to
the specific initial condition. Given this caveat, we can also in this case use Eqs.~(\ref{pot_norm1}) and~(\ref{pot_norm2}) to derive appropriate metrics for the external potential in
the way presented above.


\subsection{Potential metric for eigenstates} \label{sec:eigenstates}

For system eigenstates, Eq.~(\ref{energy_cons0}) becomes
\begin{equation} \label{energy_cons_es}
 \int\ldots\int E_{i}\abs{\psi_{i}\rbr{\mbf{r}_{1},\ldots,\mbf{r}_{N}}}^2 d\mbf{r}_{1}\ldots d\mbf{r}_{N} = E_{i}N.
\end{equation}
The norms for the external potential can then be rewritten as
\begin{align}
\int\ldots\int \abs{\rbr{E_i+c}\abs{\psi_i}^{2}}\dr_{1}\ldots d\mbf{r}_{N} = \abs{\rbr{E_i+c}N}, \label{pot_norm1es}\\
\int\abs{\rbr{E_i+c}\rho_{i}\rbr{\mbf{r}}}\dr = \abs{\rbr{E_{i}+c}N}.\label{pot_norm2es}
\end{align}
From here the metrics for the external potential become
\begin{align}
D_{v_{1}}=&\int\ldots\int\abs{\rbr{E_{1_{i}}+c_{1}}\abs{\psi_{1_{i}}}^{2}-\rbr{E_{2_{j}}+c_{2}}\abs{\psi_{2_{j}}}^{2}}\nonumber\\
&\times\dr_{1}\ldots d\mbf{r}_{N}, \label{pot_metric1es}\\
D_{v_{2}}=&\int\abs{\rbr{E_{1_{i}}+c_{1}}\rho_{1_{i}}\rbr{\mbf{r}}-\rbr{E_{2_{j}}+c_{2}}\rho_{2_{j}}\rbr{\mbf{r}}}\dr\label{pot_metric2es}.
\end{align}

\subsection{Coulomb External Potentials} \label{sec:coulomb}

Often bare Coulomb potentials are replaced by softened potentials that are finite at $r=0$. One example is the modelling of one-dimensional quantum
systems~\citep{Javanainen1988,Elliott2012}. When considering softened Coulomb potentials the external potential metrics defined above in Eqs.~(\ref{pot_metric1})
and~(\ref{pot_metric2}) are well defined. However, when the external potential has the bare Coulomb form $v=-1/r$, it diverges to $-\infty$ as $r\rightarrow 0$.
This implies that, if $\psi\rbr{\mbf{r}_{1},\ldots,\mbf{r}_{i}=0,\ldots,\mbf{r}_{N}}\neq0$ for at least one value of $i$ and $\rho\rbr{0}\neq0$, it does not seem possible
for a gauge transformation to enable the integrand of the potential norms~(\ref{pot_norm1}) and~(\ref{pot_norm2}), respectively, to be positive semi-definite everywhere.
We show below that, even in this case, the potential norms~(\ref{pot_norm1}) and~(\ref{pot_norm2}) instead remain well defined.

Let us consider the gauge transformation $v\rbr{\mbf{r}}\rightarrow v\rbr{\mbf{r}}+c$ and rewrite Eq.~(\ref{energy_cons1}) using that $\psi=\sum_{i}d_{i}\psi_{i}$, where
$\set{\psi_{i}}$ are the eigenstates of $H$, and that $H\psi_{i}=E_{i}\psi_{i}$. Equation~(\ref{energy_cons1}) then becomes
\begin{align} \label{energy_cons_es_tot}
 \int&\ldots\int\sum_{i}\rbr{E_{i}+c}\abs{d_{i}}^2\abs{\psi_{i}\rbr{\br_{1},\ldots,\br_{N}}}^2 d\mbf{r}_{1}\ldots d\mbf{r}_{N}\nonumber\\
 &=\rbr{E+c}N.
\end{align}
Equation~(\ref{energy_cons_es_tot}) shows that, as long as $\abs{E_{i}}<\infty$ for any $i$, we can choose a finite $c>0$ such that the integrand in Eq.~(\ref{energy_cons_es_tot})
is positive semi-definite everywhere, even when $v\rbr{\mbf{r}}$, as for the bare Coulomb potential, is not bounded from below.

\section{Gauge Freedom and Physical Considerations}

In Sec.~\ref{sec:metric}, we demonstrated that a gauge transformation is necessary in order to ensure that the metrics~(\ref{pot_metric1}) and~(\ref{pot_metric2}) are well defined.
The gauge must ensure that the integrands in Eqs.~(\ref{energy_cons1}) and~(\ref{energy_cons2}), respectively, are positive semi-definite everywhere, but one could make different
choices of gauge once this condition is fulfilled.

The gauge freedom we are considering reflects the fact that energies are defined up to a constant; however, energy differences have physical significance: When considering problems
where it is necessary that the (physical) difference in energy between the systems we are comparing is preserved, we must ensure that we always work in the same gauge for all systems of interest.
Hence, the constant $c$ should be the same for \emph{all} of the external potentials that we consider. In fact, from Eqs.~(\ref{pot_norm1}) and~(\ref{pot_norm2}) we see that in this
way the energy of each system is modified by the same amount, and hence the energy difference between any two systems remains unaffected. For $c$ to satisfy this condition, it must be
sufficiently large so that the integrand of Eq.~(\ref{energy_cons1}) or Eq.~(\ref{energy_cons2}) is positive semi-definite everywhere for \emph{all} of the potentials characterising the set
of systems $\set{S_{n}}$ under consideration. This condition is satisfied for any $c\geqslant\bar{c}_{1\rbr{2}}$, with $\bar{c}_{1}$ and $\bar{c}_{2}$ defined as
\begin{align}
\bar{c}_{1}\equiv\min\{&c\in\mathbb{R}\text{ s.t. }f\rbr{\mbf{r}_{1},\ldots,\mbf{r}_{N}}\geqslant 0,\nonumber \\
&\forall \set{\mbf{r}_{1},\dots,\mbf{r}_{N}}\text{ and }\forall\ S~\in\set{S_n}\},\label{cmin1}\\
\bar{c}_{2}\equiv\min\{&c\in\mathbb{R}\text{ s.t. }h\rbr{\mbf{r}}\geqslant 0,\forall\ \mbf{r}\text{ and }\forall\ S~\in\set{S_n}\},\label{cmin2}
\end{align}
for the metrics $D_{v_{1}}$ and $D_{v_{2}}$ respectively.

\section{Model Systems}

In order to assess the performance of the potential metrics $D_{v_{1}}$ and $D_{v_{2}}$ and examine the two core theorems of DFT, we will study model systems
for which we can obtain both the many-body and exact Kohn-Sham quantities with high accuracy. Since it is possible to reverse engineer the Kohn-Sham
equations exactly for systems of two electrons~\citep{Perdew1982,Laufer1986,Filippi1994}, we will study two-electron model systems, namely, Hooke's atom and the Helium atom.
Their Hamiltonians are
\begin{align}
 \hat{H}_{HA}&=\frac{1}{2}\rbr{\mbf{p}_{1}^{2}+\omega^{2}r_{1}^{2}+\mbf{p}_{2}^{2}+\omega^{2}r_{2}^{2}}+\frac{1}{\abs{\br_{1}-\br_{2}}},\\
 \hat{H}_{He}&=\frac{1}{2}\mbf{p}_{1}^{2}-\frac{Z}{r_{1}}+\frac{1}{2}\mbf{p}_{2}^{2}-\frac{Z}{r_{2}}+\frac{1}{\abs{\br_{1}-\br_{2}}}.
\end{align}
Hooke's atom can be solved exactly for particular frequencies via the method of Ref.~\citep{Taut1993}, and numerical solutions for all frequencies can
be found by the methods of Ref.~\citep{Coe2008}.

We solve the Helium atom with the variational method~\citep{Accad1971,Coe2009}.
For our purposes, we need a basis set that will allow us to obtain the ground state for any entry in the Helium isoelectronic series, i.e., 
two-electron ions with any nuclear charge $Z$. The basis set chosen is
\begin{equation} \label{Helium_basis}
 \chi_{ijk}\rbr{\br_{1},\br_{2}}=c_{ijk}N_{ijk}L_{i}^{\rbr{2}}\rbr{2Zr_{1}}L_{j}^{\rbr{2}}\rbr{2Zr_{2}}P_{k}\rbr{\cos{\theta}},
\end{equation}
with
\begin{equation}
 N_{ijk}=\sqrt{\frac{1}{\rbr{i+1}\rbr{i+2}}}\sqrt{\frac{1}{\rbr{j+1}\rbr{j+2}}}\sqrt{\frac{2k+1}{2}},
\end{equation}
where $L_{n}^{\rbr{2}}$ are the generalised Laguerre polynomials, $P_{n}$ are Legendre polynomials, and $\theta$ is the angle between $r_{1}$ and $r_{2}$.
The wavefunction for the Helium atom is then
\begin{align} \label{Helium_wave}
 \psi\rbr{\br_{1},\br_{2}}=&\frac{1}{\sqrt{8}\pi}e^{-Z\rbr{r_{1}+r_{2}}}\sum_{i,j,k}^{i+j+k\leqslant\Omega}\chi_{ijk}\rbr{\br_{1},\br_{2}},
\end{align}
where the parameter $\Omega$ controls the number of basis functions~\citep{Accad1971}.

This choice of basis combines the approaches taken by Accad \textit{et al.}~\citep{Accad1971} and Coe \textit{et al.}~\citep{Coe2009}.
It has the important advantages that, with the constants $N_{ijk}$, basis functions are orthonormal and separable in the three coordinates $\rbr{2Zr_{1},2Zr_{2},\cos{\theta}}$.
These coordinates are chosen so that the basis function with $i,j,k=0$ corresponds to the ground state of a hydrogen-like atom of charge $Z$.
This basis function always makes the largest contribution to the ground state (i.e., $c_{000}>>c_{ijk}$), particularly for large $Z$, and hence enables the ground state to converge more
rapidly with respect to the number of basis functions.

For both model systems, we will generate families of states for the metric analysis by varying a parameter in the external potentials of our systems.
For Hooke's atom, we will vary the strength of the harmonic confinement via the frequency $\omega$, and for the Helium-like atoms
we will vary the nuclear charge $Z$.

\subsection{Solving the Kohn-Sham Equations for the Model Systems} \label{calc_ks}

In order to be able to apply our metrics to quantities in the exact Kohn-Sham picture, we must be able to solve the Kohn-Sham equations exactly.
Since the exact Kohn-Sham equations must reproduce the density from the many-body picture, we can use the exact density to reverse engineer the Kohn-Sham equations.

For our model systems, the ground state is a spin singlet.
Therefore, in the Kohn-Sham picture, both electrons are described by the same Kohn-Sham orbital and, thus, are expressed in terms of the exact density as~\citep{Filippi1994}
\begin{equation} \label{KS_orbital}
 \phi_{KS}=\sqrt{\frac{\rho\ofr}{2}}.
\end{equation}
The Kohn-Sham potential follows as~\citep{Filippi1994}
\begin{equation} \label{KS_potential}
 v_{KS}\ofr=\epsilon_{KS}+\frac{1}{2}\frac{\nabla^{2}\phi_{KS}}{\phi_{KS}}.
\end{equation}
In order to obtain $v_{KS}\ofr$ from Eq.~(\ref{KS_potential}), we require the value of the Kohn-Sham eigenvalue, $\epsilon_{KS}$.
Reference~\citep{Perdew1982} demonstrated that, provided $v_{xc}\ofr\rightarrow 0$ as $\br\rightarrow\infty$,
the eigenvalue of the highest occupied Kohn-Sham state is equal to the ionisation energy of the system.

For our model systems, only one Kohn-Sham state is occupied, and thus the eigenvalues for both electrons are equal to the ionisation energy. For Hooke's atom,
when decomposed into centre-of-mass and relative motion components~\citep{Taut1993}, the centre-of-mass energy is identical to that of a one-electron
harmonic oscillator of frequency $2\omega$, so the ionisation energy is clearly equal to the relative motion energy~\citep{Laufer1986,Filippi1994}. Ionising an electron from
any entry in the Helium isoelectronic series results in a Hydrogenic atom with energy $-Z^{2}/2$ Hartrees. Therefore, the ionisation energy is found from
the difference between the Helium and the Hydrogen ground-state energies.

In order to apply our metrics to Kohn-Sham quantities, we need to consider the Hamiltonian of the whole $N$-particle Kohn-Sham system. The corresponding
Schr\"{o}dinger equation is simply the sum of the Kohn-Sham equations for each electron, so the wavefunction is formed by taking the Slater determinant of the Kohn-Sham orbitals:
\begin{align} \label{KS_slater}
 \psi_{KS}\rbr{\br_{1},\br_{2}}&=
 \begin{vmatrix}
 \phi_{KS}\rbr{\br_{1}}\uparrow_{1} & \phi_{KS}\rbr{\br_{2}}\uparrow_{2}\\
 \phi_{KS}\rbr{\br_{1}}\downarrow_{1} & \phi_{KS}\rbr{\br_{2}}\downarrow_{2}
 \end{vmatrix},\nonumber\\
 &=\phi_{KS}\rbr{\br_{1}}\phi_{KS}\rbr{\br_{2}}\rbr{\uparrow_{1}\downarrow_{2}-\downarrow_{1}\uparrow_{2}}.
\end{align}
We consider only the orbital part of the wavefunction in this paper, so the two-electron Kohn-Sham wavefunction simplifies to
%
\begin{align} \label{KS_two_e_wavefunction}
 \psi_{KS}\rbr{\br_{1},\br_{2}}&=\phi_{KS}\rbr{\br_{1}}\phi_{KS}\rbr{\br_{2}}\nonumber\\
 &=\frac{1}{2}\sqrt{\rho\rbr{\br_{1}}\rho\rbr{\br_{2}}}.
\end{align}
The potential for the two Kohn-Sham electrons' Hamiltonian is given by the sum of the single-particle Kohn-Sham potentials,
\begin{equation} \label{KS_two_e_potential}
 V_{KS}\rbr{\br_{1},\br_{2}}=v_{KS}\rbr{\br_{1}}+v_{KS}\rbr{\br_{2}}.
 \end{equation}
We will apply our metrics to these two-electron Kohn-Sham quantities. Equation~(\ref{KS_two_e_wavefunction}) shows that for a Kohn-Sham system
the metrics $D_{v_{1}}$ and $D_{v_{2}}$ will, in general, take on different values.

\begin{figure*}
 \begin{center}
 \includegraphics[width=\textwidth]{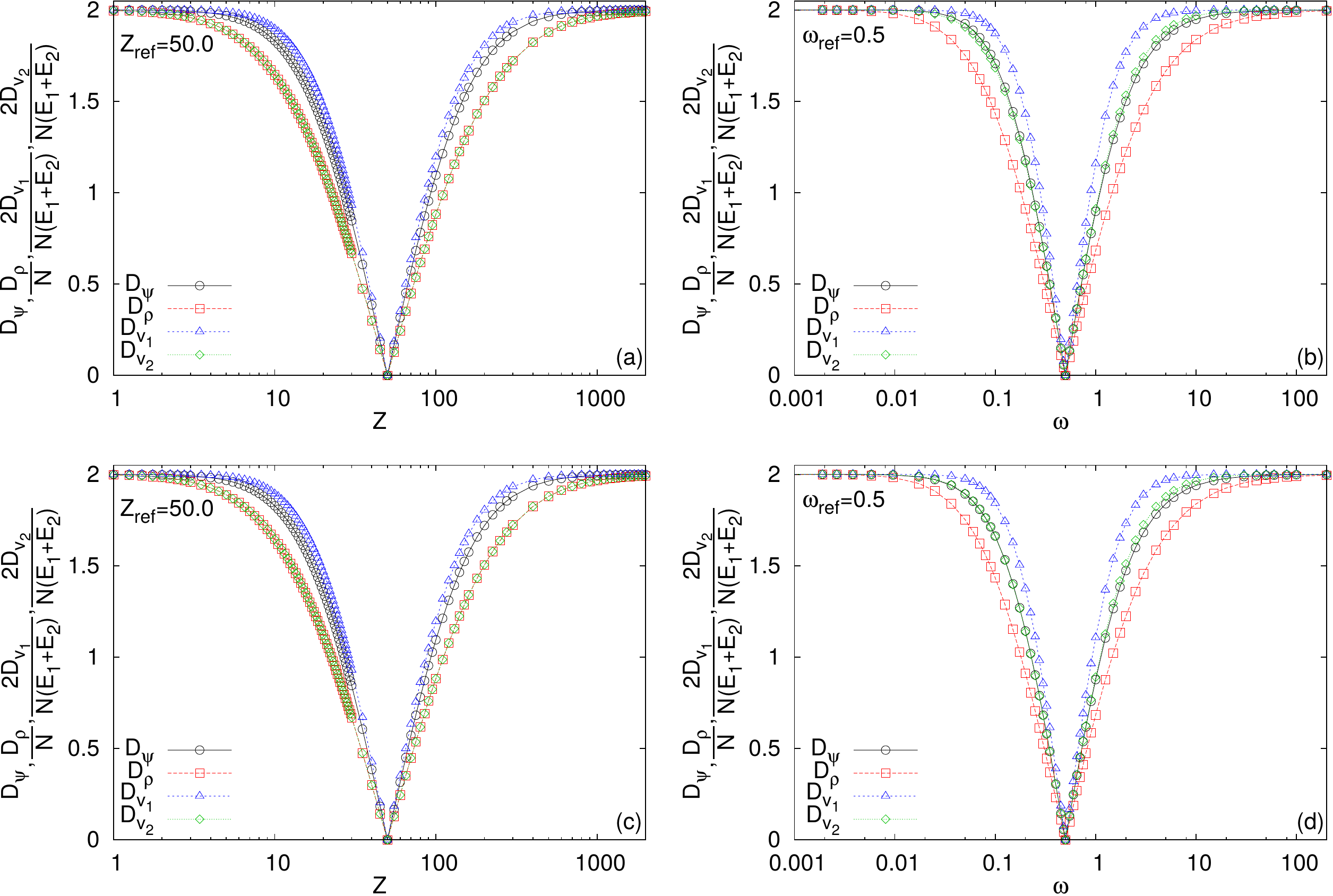}
 \caption{(Color online) The wavefunction, density, and potential distances for many-body systems [(a) and (b)] and Kohn-Sham systems [(c) and (d)] are plotted against the
 nuclear charge for Helium-like atoms (left), and against the confinement frequency for Hooke's atom (right). For Helium-like atoms the reference state is $Z=50.0$,
 and for Hooke's atom the reference state is $\omega=0.5$. All of the metrics are scaled such that their maximum value is $2$.}
 \label{compare_metrics}
 \end{center}
\end{figure*}

\section{Comparison of Metrics for Characterising Quantum Systems}

\begin{figure*}[t]
 \includegraphics[width=\textwidth]{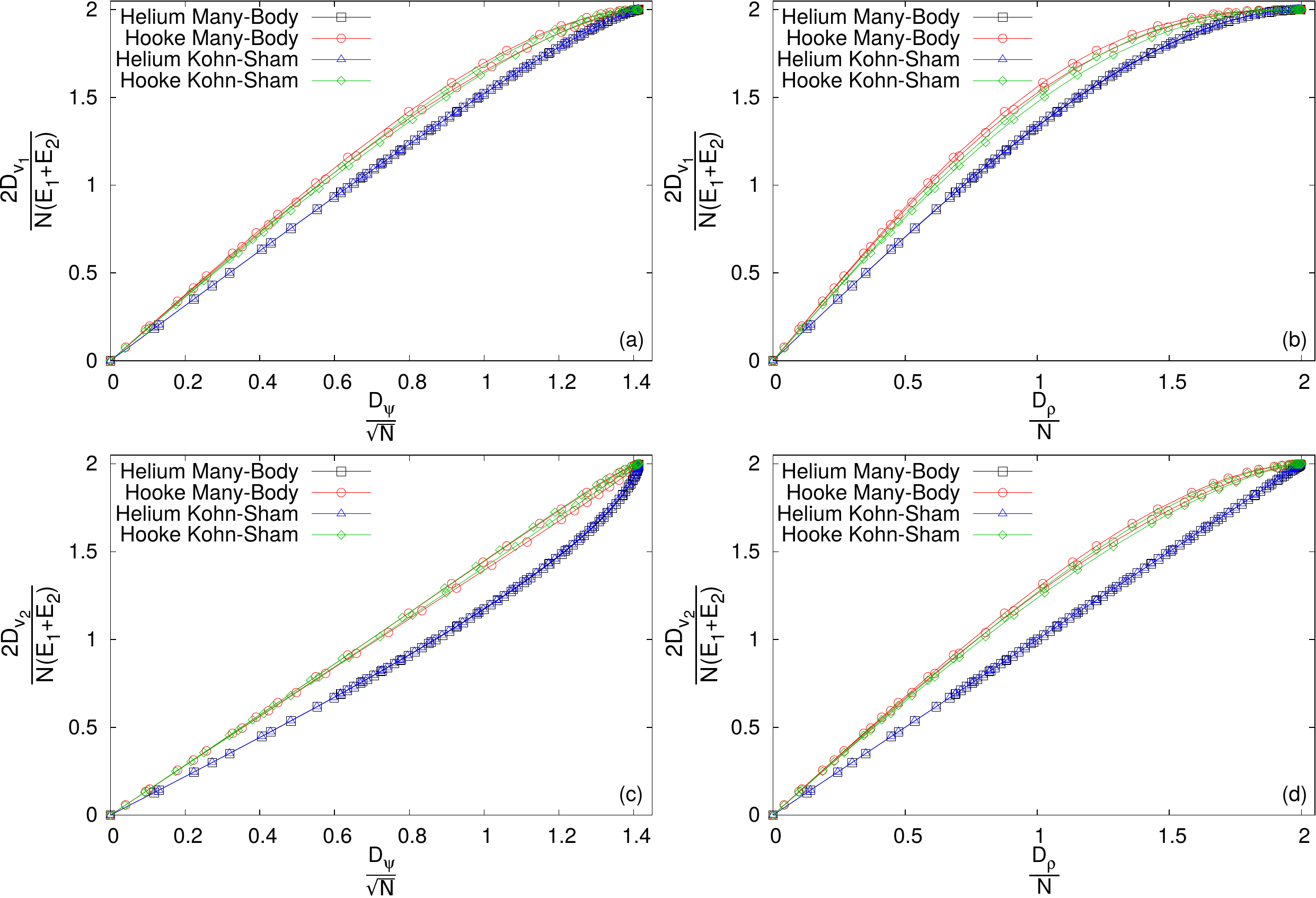}
 \caption{(Color online) Plots of rescaled potential distance $2D_{v_{1}}/[N(E_1+E_2)]$ (top) and $2D_{v_{2}}/[N(E_1+E_2)]$ (bottom) against rescaled wavefunction
 distance $D_{\psi}/\sqrt{N}$ [(a) and (c)] and against rescaled density distance $D_{\rho}/N$ [(b) and (d)]. We have plotted both the many-body and related Kohn-Sham
 systems for Helium-like atoms and Hooke's atom. In each panel we consider families of systems characterised by increasing and decreasing parameters starting from the
 reference state ($Z=50.0$ for Helium-like atoms, $\omega=0.5$ for Hooke's atom). The parameter ranges are $1.0<Z<2000.0$ for Helium-like atoms, and
 $2.6\times10^{-8}<\omega<1000.0$ for Hooke's atom.}
 \label{hk_dv}
\end{figure*}

Within the metric space approach to quantum mechanics, we now have metrics for wavefunctions, densities, and potentials.
For systems subject only to scalar potentials and with a given many-body interaction, these quantities, taken together, fully characterise a many-body system.
We are then, in principle, in the position of \textit{quantitatively} answering the following questions. Are two many-body systems close to each other in the Hilbert space? Could two many-body
systems be close to each other with respect to some of these quantities but far away for others? We will address these questions, at least for the systems at hand and
with a focus on DFT, in the rest of the paper: Apart from the general interest, these questions have practical implications, for example when considering how closely
quantum information processes reproduce the desired result~\cite{Nielsen2000} or assessing the effectiveness of convergence loops in codes aiming to 
determine numerically accurate properties of systems, such as DFT codes. 

When considering ground states, thanks to the Hohenberg-Kohn theorem,
any among the density, wavefunction, and external potential are equally appropriate for characterising quantum systems subject to external scalar potentials.
Therefore, it is worthwhile to make a comparison between the information given by each of the corresponding metrics.

Figure~\ref{compare_metrics} shows the values of the wavefunction, density, and both potential metrics plotted against the parameter values for both of our
model systems and considering both many-body (top panels) and Kohn-Sham (bottom panels) quantities. The distances are calculated with respect to a reference state,
$Z=50.0$ for the Helium-like atoms and $\omega=0.5$ for Hooke's atom, and are all scaled to have a maximum value of $2$ for ease of comparison.
We can immediately observe that all of the metrics follow broadly the same trend, increasing monotonically from the reference to their maximum value.
The curves for both increasing and decreasing values of the parameters incorporate a region of rapidly increasing distance for parameter values close to the reference,
a region where the distance asymptotically approaches its maximum for parameter values far from the reference, along with a transition region in between,
where the largest differences between metrics are observed. The crucial difference between the four metrics, however, is how the metrics converge to the maximum value.
Figure~\ref{compare_metrics} shows that, as we depart from the reference, the potential metric $D_{v_{1}}$ is the fastest to converge to its maximum, followed
by the wavefunction metric, with the density metric being the slowest to converge. The behaviour of the metric $D_{v_{2}}$ is different for the two systems that we study.
We firstly note the metric $D_{v_{2}}$ takes on different values for many-body and Kohn-Sham systems because, although they share the same density, many-body and related Kohn-Sham
systems have different energies in general. For Helium-like atoms, this metric strongly follows the trend of the density metric for both many-body and Kohn-Sham quantities.
However, when considering Hooke's atom, the potential metric $D_{v_{2}}$ is similar in value to the wavefunction metric, albeit slightly greater for frequencies greater than the reference.
These results suggest that, when comparing systems that are significantly different from one another, the density metric is the most useful tool for analysis, as it is capable of
providing non-trivial information over a wider range of parameter space than the metrics for wavefunctions and potentials. When comparing systems that are
relatively close to one another, all four metrics provide useful information to quantitatively characterise the differences between the systems.

With regard to practical calculations, the density metric $D_{\rho}$, along with the potential metric $D_{v_{2}}$, has another
significant advantage in that, in general, it is considerably easier to calculate than the metrics $D_{\psi}$ and $D_{v_{1}}$.
The metrics $D_{\rho}$ and $D_{v_{2}}$, in fact, need only be integrated over three degrees of freedom, compared to $3N$ degrees of freedom for the other two metrics.
Also we can calculate the density metric from both the many-body and Kohn-Sham systems, since, unlike for wavefunctions and potentials, the Kohn-Sham system will, in principle,
provide the exact many-body density.

\section{Mappings relevant to the Hohenberg-Kohn Theorem}

\begin{figure}
 \includegraphics[width=\columnwidth]{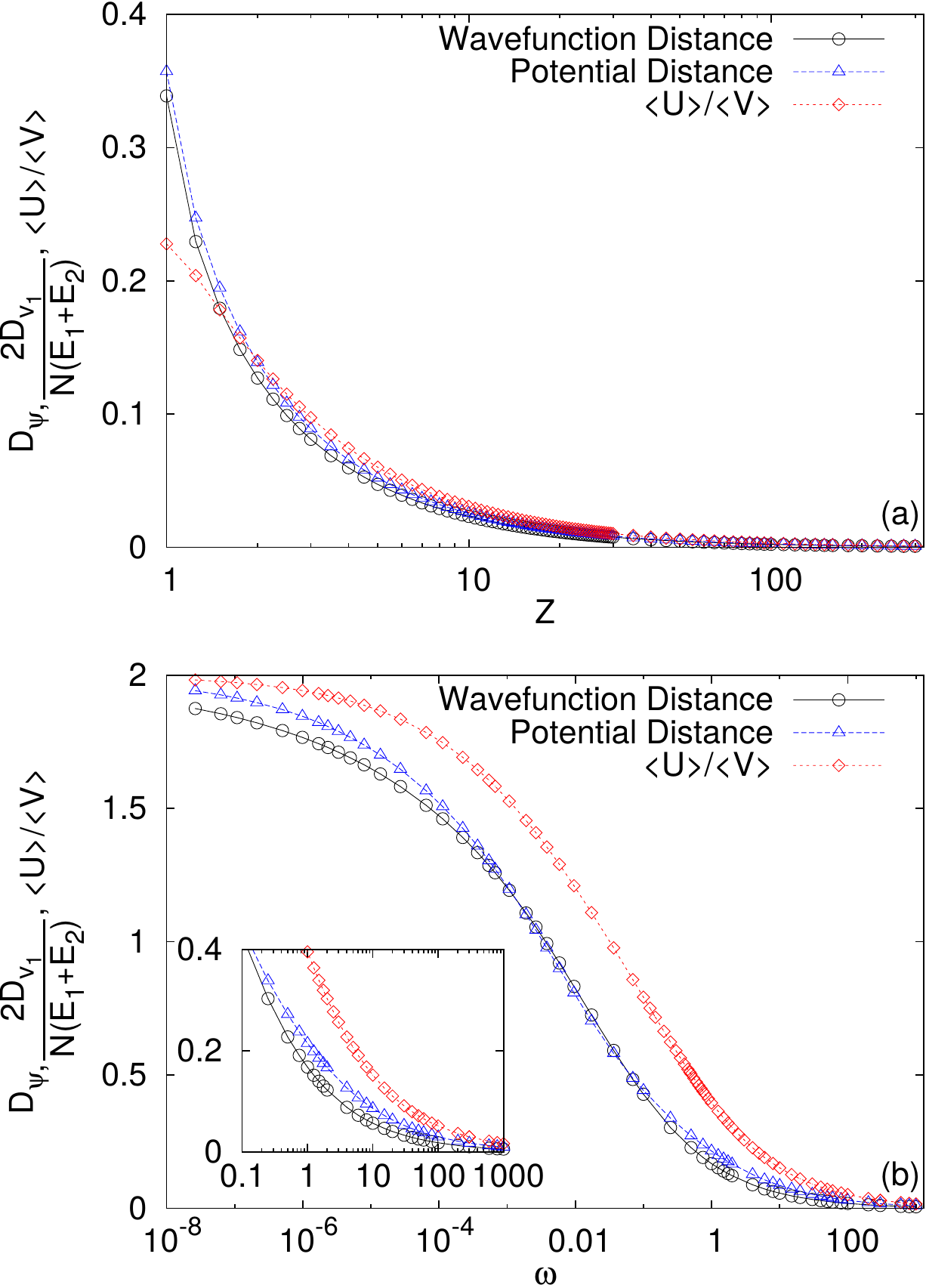}
 \caption{(Color online) For (a) Helium-like atoms and (b) Hooke's atom, the distances between many-body
 and Kohn-Sham wavefunctions, and between many-body and Kohn-Sham potentials, are plotted against the parameter values. In addition, the ratio of the expectation of
 the electron-electron interaction to the many-body external potential energy is plotted and shown to follow a similar trend to the metrics.
 In the inset, we focus on Hooke's atom in the regime of distances covered by the Helium-like atoms.}
 \label{mb_ks}
\end{figure}

In Ref.~\citep{D'Amico2011} it was shown that the mapping between wavefunctions and densities in the Hohenberg-Kohn theorem [Eq.~(\ref{HK_map})] is a mapping between metric spaces; by examining it
in this light several features were found. In this paper, we have shown that all of the relationships in Eq.~(\ref{HK_map}) are mappings between metric spaces:
Using various families of states for each of our model systems, we will now look at the other relationships within the Hohenberg-Kohn theorem. 
We choose a reference state for each family of systems. We then calculate the distance between each member of the family and the reference state, for densities, wavefunctions, and potentials.

In Fig.~\ref{hk_dv} we plot the potential metrics $D_{v_{1}}$ and $D_{v_{2}}$, respectively, against the wavefunction (left-hand panels) and density (right-hand panels) metrics
for both interacting systems and their related Kohn-Sham systems and for increasing and decreasing parameters.
In this way we compare for each plot eight different families of states as well as the behaviour of the many-body systems with
respect to the non-interacting Kohn-Sham systems. The rescaling of the metrics has been chosen such that the dependence on the particle number is removed and that these figures are directly comparable
to Fig.~2 of Ref.~\citep{D'Amico2011}, where corresponding plots for $D_{\psi}$ versus $D_{\rho}$ for Helium-like and Hooke's atoms were considered.

Considering our plots, we observe many features in common with the relationship between wavefunction and density metrics of Ref.~\citep{D'Amico2011}: The relationships between the potential distances
and the other distances are monotonic, with nearby wavefunctions and nearby densities mapped onto nearby potentials and distant wavefunctions and distant densities mapped onto distant potentials.
The curves for increasing parameters and decreasing parameters within each of the four systems (Hooke's many-body, Hooke's Kohn-Sham, Helium-like many-body, Helium-like Kohn-Sham) are also seen to
overlap, or almost overlap, with one another. Finally, all curves have an extended region (up to and including intermediate potential distances) where the relationship between potential and the other
distances is linear or almost linear. Interestingly, depending on the potential distance and the system considered, we observe that this linear region can cover the entire parameter range; see
Figs.~\ref{hk_dv}(a), \ref{hk_dv}(c), and~\ref{hk_dv}(d). With the exception of Fig.~\ref{hk_dv}(c), we notice that the curves have opposite convexity at large distances with respect to Fig.~2
of Ref.~\citep{D'Amico2011}, which suggests that, in general, the potential distance is more likely to converge to its maximum faster than wavefunction or density distances;
hence, in general, it is less effective in distinguishing far-away systems (compare also with Fig.~\ref{compare_metrics}).

In Ref.~\citep{D'Amico2011} a hint to universality was observed for the mapping between wavefunction and density distances; when looking at the potential versus wavefunction or density
distances we note that the mapping from each many-body system is very close to the one from the corresponding exact Kohn-Sham system.
This mapping is closer for Helium-like atoms compared to Hooke's Atom; this is because we are always in a weak-correlation regime for Helium-like atoms, while we consider
both strong and weak correlation regimes for Hooke's Atom (see Fig.~\ref{mb_ks}). However, the mapping is less close when comparing the behaviour of Hooke's with respect to Helium-like atoms,
and particularly so for the $D_{v_2}$ distance, for which the convexity of the corresponding curves at large distances may be opposite [compare curves for the two Kohn-Sham systems in Fig.~\ref{hk_dv}(c)].

\section{Quantitative analysis of the Kohn-Sham Scheme}

We will now consider the distance between wavefunctions and potentials of many-body systems, and the ones used to describe the corresponding Kohn-Sham
systems~\footnote{For densities, it is required that $D_{\rho}\rbr{\rho_{MB},\rho_{KS}}\equiv0$.}, and study how these distances change throughout the parameter range.
This allows us to provide a \textit{quantitative} description of the differences between the many-body and exact Kohn-Sham descriptions of quantum systems.
Although there is no promise from DFT for the many-body wavefunction to be reproduced by the Kohn-Sham ground-state wavefunction, the latter is commonly used as an
approximation to the former in various contexts, such as linear response calculations in time-dependent DFT and some magnetic-system calculations, even if the regime
of validity of this approximation has not been properly established. It is therefore of interest to quantitatively determine how good this approximation is.

In Fig.~\ref{mb_ks}, the distances between many-body and Kohn-Sham wavefunctions and potentials are plotted for a range of parameter values. For potentials, we use here the
metric $D_{v_{1}}$, since Eq.~(\ref{pot_metric2es}) shows that, in this case, the metric $D_{v_{2}}$ will yield only the difference in the energy of the two systems. We first
observe that the wavefunction and potential distances, when rescaled to the same maximum value, always take approximately the same value throughout the parameter range explored
for both systems. This demonstrates that the two metrics provide a consistent measure of how the many-body description differs from the Kohn-Sham description of our systems.

For both systems we have also plotted the ratio of the Coulomb energy to the external
potential energy for the many-body systems. This ratio can be seen to follow broadly the the same trend as the metrics. This is an important observation as it provides further
confirmation that the metrics derived from the metric space approach to quantum mechanics provide a physically relevant comparison of quantum mechanical
functions. It also shows that, alongside the two metrics and at least for the systems considered, this ratio is a useful indicator of how much the many-body and Kohn-Sham descriptions of the system
differ from one another.

If we consider as a good performance indicator that the distance between the many-body and Kohn-Sham wavefunctions is up to 10\% of the maximum distance
[i.e., $D_{\psi}\rbr{\psi_{MB},\psi_{KS}}<0.2$], then we see that for all families of systems the Kohn-Sham wavefunction is indeed a good approximation for a relatively large
range of parameters, for $Z>1.5$ for the Helium isoelectronic series and $\omega>1.25$ for Hooke's atom.

For Helium-like atoms, even at $Z=1$, the maximum difference between the many-body and Kohn-Sham systems is just 17.5\%.
For these systems, the external potential always dominates over the Coulomb interaction between the electrons, and
we observe that the distance between the potentials is always larger than the distance between the wavefunctions.
For Hooke's atom, for small and large values of $\omega$, we observe that the value of the potential metric is greater than that of the wavefunction metric, while,
in the region where the ratio $\an{U}/\an{V}$ is approximately unity, the wavefunction metric takes a larger value than the potential metric.

In the inset of Fig.~\ref{mb_ks}, we show the large $\omega$ behaviour of our metrics for Hooke's atom, which can be seen for Helium-like atoms in Fig.~\ref{mb_ks}(a).
In this regime, both metrics and the ratio $\an{U}/\an{V}$ all tend to zero. This behaviour can be understood by considering the limit of the
quantities of interest in the regime where the external potential strongly dominates over the Coulomb interaction. The Kohn-Sham external potential is the sum of the external
potential used to describe the many-body system, the Hartree potential, and the exchange-correlation potential; in this regime, $V_{KS}\approx V_{ext}$, and
hence $D_{v_{1}}\rbr{V_{KS},V_{ext}}\approx0$. Likewise, the many-body wavefunction approaches a non-interacting wavefunction which coincides with the Kohn-Sham wavefunction;
hence, $D_{\psi}\rbr{\psi_{MB},\psi_{KS}}\approx0$.

Physically, the wavefunction and potential distances between many-body and Kohn-Sham systems can be interpreted as
a measure of specific electron-electron interaction effects. The Kohn-Sham wavefunction is the product of single-particle states;
hence, the wavefunction distance can be interpreted as a measure of the features of the many-body wavefunction that go beyond
single-particle approximations. In this respect this distance is a measure of correlation effects, which cannot be captured by mean-field-type approximations.
For potentials, the value of the metric $D_{v_{1}}\rbr{V_{ext},V_{KS}}$ can be interpreted as measuring the contribution of the Hartree and exchange-correlation
potentials to the Kohn-Sham potential.

\section{Conclusion}

The aim of this paper was to derive a metric for external potentials, which is motivated by their role in the Hohenberg-Kohn theorem, and more generally the crucial role external potentials play
in modelling quantum systems. This metric complements the density and wavefunction metrics, providing us with metrics for each of the fundamental quantities of DFT. The tools we now have at our
disposal have enabled us to take our metric analysis in other directions, such as the quantitative analysis of the Kohn-Sham scheme.
In particular, since the density of Kohn-Sham and many-body interacting systems are the same, the potential metric is able to provide a meaningful insight into the Kohn-Sham scheme that the density metric cannot.

By considering the conservation of energy and applying the metric space approach to quantum mechanics to it, we have derived two ``natural'' metrics for external potentials.
These metrics can be applied to electronic systems subject to any physical scalar potential (including unbounded potentials such as Coulomb interactions), in eigenstates or out of equilibrium.
We also showed how to extend our analysis to derive the potential metrics for systems incorporating both electronic and nuclear effects.
This analysis can be straightforwardly extended to even more complex systems.
We have also considered the effects of the gauge freedom of potentials and shown which conditions the metrics should satisfy to remain well defined when the preservation of
relative energy differences are important to the problem considered. As for all metrics derived within the metric space approach to quantum mechanics, our potential metrics are
characterised by well-defined maximum values, which makes it possible to compare quantitatively the behaviours of very different systems.

Physical systems subject to scalar potentials are defined through their external potentials, densities and wavefunctions: Here we have analysed in detail eight families of systems,
all in their ground states, so that these quantities are subject to a one-to-one mapping through the Hohenberg-Kohn theorem, the pillar of Density Functional Theory. These families
are defined by increasing and decreasing parameters with respect to reference systems for the interacting Helium isoelectronic series, the interacting Hooke's atom with varying
confinement strength, and the two corresponding families of non-interacting exact Kohn-Sham systems. When comparing the performances of the metrics, we found that
they converged onto their maximum values at different rates, with the potential metric $D_{v_{1}}$ converging first, followed by the wavefunction metric, and finally
by the density metric, with the behaviour of the potential metric $D_{v_{2}}$ depending on the system studied. This strengthens the findings in Ref.~\citep{D'Amico2011} that the density
is the best quantity to differentiate between distant systems. Importantly, however, we find that, in general, two systems close to (or distant from) each other with respect to the
metric for one physical quantity remain so with respect to the metrics for all physical quantities.

In the context of the Hohenberg-Kohn theorem, in Ref.~\citep{D'Amico2011} it was found that in metric spaces the mapping between wavefunctions and densities was monotonic, and incorporated
a (quasi) linear mapping between small and between intermediate distances. When examining in metric space the relationships of the external potential with wavefunctions and densities
in the Hohenberg-Kohn theorem, we find once more surprisingly simple mappings and with a similar behaviour, with some curves showing an even greater range of linearity than the
wavefunction-density mapping. These results are evidence of the deep connection between the quantities involved in the Hohenberg-Kohn theorem.
However, while the interacting and related exact Kohn-Sham systems have almost identical behaviour, there are differences, especially at intermediate to large
distance regions between Hooke's and Helium-like families, as opposed to Ref.~\citep{D'Amico2011}.

We looked at the distance between many-body and Kohn-Sham quantities for both wavefunctions and external potentials, gaining quantitative insight into when, and by how much,
the many-body and Kohn-Sham systems differ from one another. We showed that, when rescaled to the same maximum distance, wavefunctions and potentials provide a consistent picture,
since they yield approximately the same distance values throughout all the parameter ranges considered. We also found that the two metrics followed the same qualitative trend as the
ratio of Coulomb to external potential energies. The Kohn-Sham wavefunction has been used as an approximation to the many-body wavefunction, even if there is no promise of good behaviour,
in this respect, from density functional theory. Our metrics allowed us to explore this approximation \textit{quantitatively}, at least for the systems at hand. For these systems we prove that
the Kohn-Sham wavefunction indeed represents a well-behaved approximation which provides good quantitative results (10\% maximum error) for a relatively large range of the parameters explored.

\begin{acknowledgments}
We acknowledge fruitful discussions with E.K.U.~Gross.
P.M.S. acknowledges support from EPSRC.
P.M.S. and I.D. acknowledge support from Royal Society Grant NA140436 and CNPq Grant: PVE--Processo: 401414/2014-0.
All data published during this research are available by request from the University of York Data Catalogue 10.15124/dc3868e7-38eb-4ef0-b97c-210773f2251c
\end{acknowledgments}

\appendix*

\section{External potential metrics for systems comprising electrons and nuclei}
In this appendix we will generalise the external potential metrics $D_{v_{1}}$ and $D_{v_{2}}$ to systems comprising both electrons and nuclei.
We define the sum of the electrons and nuclei numbers $N_e+N_n\equiv N$, and consider the Hamiltonian
\begin{equation} \label{nuc_hamiltonian}
\hat{H}= -\sum_{i=1}^{N}\frac{1}{2}\nabla_{i}^{2} + \sum_{j<i}^{N} U\rbr{\br_{i},\br_{j}} + \sum_{i=1}^{N_e} v_e\rbr{\mbf{r}_{i}}+\sum_{i=1}^{N_n} v_n\rbr{\mbf{r}_{i}},
\end{equation}
where $V=\sum_{i=1}^{N_e} v_e\rbr{\mbf{r}_{i}}+\sum_{i=1}^{N_n} v_n\rbr{\mbf{r}_{i}}$ is the external potential acting on the electrons and nuclei (e.g., from an applied electric field)
and $\sum_{j<i}^{N} U\rbr{\br_{i},\br_{j}}$ is a shorthand for
\begin{align}
\sum_{j<i}^{N_e+N_n}U\rbr{\br_{i},\br_{j}}\equiv&\sum_{j<i}^{N_e} U_e\rbr{\br_{i},\br_{j}} + \sum_{\substack{i=N_e+1,\\j<i}}^{N_e+N_n} U_n\rbr{\br_{i},\br_{j}}\nonumber \\
&+\sum_{i=1}^{N_e}\sum_{j= N_e+1}^{N_e+N_n} U_{e-n}\rbr{\br_{i},\br_{j}} \label{total_pot_e_n}
\end{align}
and contains the electron-electron, nuclear-nuclear, and electron-nuclear interactions, respectively. The system state is
$\psi\rbr{\br_{1},\ldots,\br_{N_e},\br_{N_e+1},\ldots,\br_{N_e+N_n}}$, where we have followed Ref.~\citep{D'Amico2011}, and normalised the many-body wavefunction to
the total particle number $N\equiv N_e+N_n$. Without loss of generality, we have positioned the electron coordinates before the nuclear coordinates.

\onecolumngrid
\subsection{Generalisation of $D_{v_{1}}$ to an electron-nuclear system}
The Hamiltonian expectation value is
\begin{align} \label{energy_cons_e_n}
\int&\ldots\int\psi^{*}\rbr{\br_{1},\ldots,\br_{N_e},\br_{N_e+1},\ldots,\br_{N_e+N_n}}\hat{H}\psi\rbr{\br_{1},\ldots,\br_{N_e},\br_{N_e+1},\ldots,\br_{N_e+N_n}} d\mbf{r}_{1}\ldots d\mbf{r}_{N_e+N_n} \nonumber\\
=&\int\ldots\int \set{-\sum_{i=1}^{N}\frac{1}{2}\psi^{*}\nabla_{i}^{2} \psi+\sbr{\sum_{j<i}^{N_e} U_e\rbr{\br_{i},\br_{j}} +\sum_{i=N_e+1,j<i}^{N_e+N_n} U_n\rbr{\br_{i},\br_{j}}}\abs{\psi}^2 \right.\nonumber\\
&+\left.\sum_{i=1}^{N_e}\sum_{j=N_e+1}^{N_e+N_n} U_{e-n}\rbr{\br_{i},\br_{j}}\abs{\psi}^2+\sbr{\sum_{i=1}^{N_e} v_e\rbr{\mbf{r}_{i}}+\sum_{i=1}^{N_n} v_n\rbr{\mbf{r}_{i}}}\abs{\psi}^2}d\mbf{r}_{1}\ldots d\mbf{r}_{N_e+N_n}=E(N_e+N_n) = EN.
\end{align}
Following a procedure similar to the one used to derive Eq.~(\ref{pot_norm1}) we can write
\begin{equation}\label{pot_int_e_n}
\int\ldots\int\sbr{F\rbr{\br_{1},\ldots,\br_{N}}+\sum_{i=1}^{N}c\abs{\psi}^2}d\mbf{r}_{1}\ldots d\mbf{r}_{N}=\rbr{E+c}N,
\end{equation}
where $F\rbr{\br_{1},\ldots,\br_{N}}$ is the integrand of (\ref{energy_cons_e_n}) and $c$ is the positive constant from the gauge transformation $v_{e(n)}\ofr\to v_{e(n)}\ofr+c$.
While the kinetic term [after applying Eq.~(\ref{ke_relation})] and the terms containing the electron-electron and the nuclear-nuclear interactions are positive definite,
this gauge transformation is necessary to ensure the the sum of the electron-nuclear and external potential terms in (\ref{energy_cons_e_n}) is also positive definite.
By using that $\psi=\sum_i d_i\psi_i$ with $\set{\psi_i}$ the set of orthogonal eigenstates such that $H\psi_i=E_i \psi_i$, Eq.~(\ref{pot_int_e_n}) can be rewritten as
\begin{equation}
\int\ldots\int \sum_i\rbr{E_i+c}\abs{d_i}^2\abs{\psi_i}^2 d\mbf{r}_{1}\ldots d\mbf{r}_{N}=\rbr{E+c}N.
\end{equation}
As was the case with Eq.~(\ref{energy_cons_es_tot}), this equation proves that, provided that $\abs{E_i}<\infty$ for all $i$, it is possible to find a value of $c$ such
that the integrand of (\ref{pot_norm1_e_n}) becomes positive definite. With this choice of $c$ we can write
\begin{equation}\label{pot_norm1_e_n}
 \int\ldots\int \abs{F\rbr{\br_{1},\ldots,\br_{N}}+ \sum_{i=1}^{N}c\abs{\psi}^2} d\mbf{r}_{1}\ldots d\mbf{r}_{N}=\abs{\rbr{E+c}N},
\end{equation}
which is the analogue of Eq.~(\ref{pot_norm1}) for the Hamiltonian~(\ref{nuc_hamiltonian}) and represents a well-defined $L^1$ norm when extended to the appropriate set~\citep{Sharp2014}.
From this, following the metric space approach to quantum mechanics~\citep{Sharp2014}, we derive the generalisation of $D_{v_1}$ to the external potential
$\sum_{i=1}^{N_e} v\rbr{\mbf{r}_{i}}+\sum_{i=1}^{N_n} v\rbr{\mbf{r}_{i}}$, which reads
\begin{align}
D_{v_{1},e-n}=&\int\ldots\int \abs{F_1\rbr{\br_{1},\ldots,\br_{N}}+ \sum_{i=1}^{N}c\abs{\psi_1}^2-F_2\rbr{\br_{1},\ldots,\br_{N}}- \sum_{i=1}^{N}c\abs{\psi_2}^2} d\mbf{r}_{1}\ldots d\mbf{r}_{N}\label{pot_metric1_e_n},\\
=&\int\ldots\int \abs{f_{1,e-n}-f_{2,e-n}}d\mbf{r}_{1}\ldots d\mbf{r}_{N},
\end{align}
where
\begin{equation}
f_{i,e-n}\rbr{\br_{1},\ldots,\br_{N}}=F_i\rbr{\br_{1},\ldots,\br_{N}}+ \sum_{j=1}^{N}c\abs{\psi_i}^2.
\end{equation}
In a similar way the metric $D_{v_1}$ can be generalised to measure the distance between systems containing an arbitrary number of sets of different particles $p_a,p_b\ldots,p_m$
(e.g. systems which include electrons and various ionic species), as long as the number of corresponding particles is identical for both systems,
i.e., $N_{a_{1}}\equiv N_{a_{2}}, N_{b_{1}}\equiv N_{b_{2}},$ etc.

\subsection{Generalisation of $D_{v_{2}}$ to an electron-nuclear system}

The system wavefunction $\psi\rbr{\br_{1},\ldots,\br_{N_e},\br_{N_e+1},\ldots,\br_{N_e+N_n}}$ is antisymmetric with respect to electron-electron exchange,
and either symmetric or antisymmetric with respect to nuclear-nuclear exchange depending on whether the nuclei are bosons or fermions, respectively.
By using these properties we can rewrite Eq.~(\ref{energy_cons_e_n}) as
\begin{align} \label{pot_norm2.1_e_n}
N&\set{\int_V d\mbf{r}_{e}\an{\tau_e\rbr{\br_{e};\br_{N_e+1},\ldots,\br_{N_e+N_n}}}_n \right.+\int_V d\mbf{r}_{n}\an{\tau_n\rbr{\br_{n};\br_{1},\ldots,\br_{N_e}}}_e+\int_V d\mbf{r}_{e_{1}}\frac{1}{2}\int_V d\mbf{r}_{e_{2}}U_e\rbr{\br_{e_{1}},\br_{e_{2}}}\nonumber\\
&\quad\times\an{g_e\rbr{\br_{e_{1}},\br_{e_{2}};\br_{N_e+1},\ldots,\br_{N_e+N_n}}}_n+\int_V d\mbf{r}_{n_{1}}\frac{1}{2}\int_V d\mbf{r}_{n_{2}}U_n\rbr{\br_{n_{1}},\br_{n_{2}}}\an{g_n\rbr{\br_{n_{1}},\br_{n_{2}};\br_{1},\ldots,\br_{N_e}}}_e   \nonumber\\
&+\int_V d\mbf{r}_{e}\int_V d\mbf{r}_{n}U_{e-n}\rbr{\br_{e},\br_{n}}g_{e-n}\rbr{\br_{e},\br_{n}}+\int_V d\mbf{r}_{e}v_{e}\rbr{\br_{e}}\an{\rho_e\rbr{\br_{e};\br_{N_e+1},\ldots,\br_{N_e+N_n}}}_n \nonumber\\
&+\left. \int_V d\mbf{r}_{n}v_{n}\rbr{\br_{n}}\an{\rho_n\rbr{\br_{n};\br_{1},\ldots,\br_{N_e}}}_e}=EN,
\end{align}
where
\begin{align}
&\an{\tau_e\rbr{\br_{e};\br_{N_e+1},\ldots,\br_{N_e+N_n}}}_n\equiv\int  d\br_{N_e+1},\ldots,d\br_{N_e+N_n} \sbr{\frac{N_e}{2N}\int \abs{\nabla_{i}\psi}^2 d\br_{2},\ldots,d\br_{N_e}}, 
\end{align}
\begin{align}
&\an{g_e\rbr{\br_{e_{1}},\br_{e_{2}};\br_{N_e+1},\ldots,\br_{N_e+N_n}}}_n\equiv\int d\br_{N_e+1},\ldots,d\br_{N_e+N_n} \sbr{\frac{N_e\rbr{N_e-1}}{N}\int \abs{\psi}^2 d\br_{3},\ldots,d\br_{N_e}}, 
\end{align}
\begin{align}
&\an{\rho_e\rbr{\br_{e};\br_{N_e+1},\ldots,\br_{N_e+N_n}}}_n\equiv\int d\br_{N_e+1},\ldots,d\br_{N_e+N_n} \sbr{\frac{N_e}{N}\int \abs{\psi}^2 d\br_{2},\ldots,d\br_{N_e}}.
\end{align}
It can be seen that the terms in square brackets in Eqs.~(A9)-(A11) correspond to the definitions of the analogous quantities for electron-only systems in Eqs.~(\ref{kinetic_density})-(\ref{1_part_density}).
The corresponding nuclear functions are obtained by interchanging in the three equations above the sets of electron and nuclear coordinates and the ``$e$'' and ``$n$'' indices, and
\begin{equation}
g_{e-n}\rbr{\br_{e},\br_{n}}\equiv\frac{N_e N_n}{N}\int d\br_{2},\ldots,d\br_{N_e}\int \abs{\psi}^2 d\br_{N_e+2},\ldots,d\br_{N_e+N_n}.
\end{equation}

We then note that (i) all integrations in (\ref{pot_norm2.1_e_n}) are over the same volume, (ii) the integrands of the first four terms are positive definite,
(iii) the integrand of the fifth term is negative, and (iv) the integrands of the sixth and seventh terms have no defined sign.
By using (i) and a gauge transformation for $v_e\rbr{\br}$ and $v_n\rbr{\br}$, we can write (\ref{pot_norm2.1_e_n}) as
\begin{align} \label{pot_norm2.2_e_n}
\int_V& N\set{\an{\tau_e\rbr{\br;\br_{N_e+1},\ldots,\br_{N_e+N_n}}}_n+\an{\tau_n\rbr{\br;\br_{1},\ldots,\br_{N_e}}}_e \right.\nonumber\\
&+\frac{1}{2}\int_V d\mbf{r'}\sbr{U_e\rbr{\br,\br'}\an{g_e\rbr{\br,\br';\br_{N_e+1},\ldots,\br_{N_e+N_n}}}_n+U_n\rbr{\br,\br'}\an{g_n\rbr{\br,\br';\br_{1},\ldots,\br_{N_e}}}_e} \nonumber\\
&+\int_V d\mbf{r}'U_{e-n}\rbr{\br,\br'}g_{e-n}\rbr{\br,\br'}+\rbr{v_{e}\rbr{\br}+c}\an{\rho_e\rbr{\br;\br_{N_e+1},\ldots,\br_{N_e+N_n}}}_n \nonumber\\
&+\left.\rbr{v_{n}\rbr{\br}+c}\an{\rho_n\rbr{\br;\br_{1},\ldots,\br_{N_e}}}_e}d\mbf{r} =  \rbr{E+c}N,
\end{align}
where $c\geqslant 0$ is chosen such that the sum of the last three terms of the overall integrand is always positive. In this way the overall integrand in (\ref{pot_norm2.2_e_n})
is positive definite and, following the metric space approach to quantum mechanics~\citep{Sharp2014}, we can write the $L^1$ norm 
\begin{align} \label{pot_norm2.3_e_n}
\int_V& N\abs{\an{\tau_e\rbr{\br;\br_{N_e+1},\ldots,\br_{N_e+N_n}}}_n+\an{\tau_n\rbr{\br;\br_{1},\ldots,\br_{N_e}}}_e \right.\nonumber\\
&+\frac{1}{2}\int_V d\mbf{r'}\sbr{U_e\rbr{\br,\br'}\an{g_e\rbr{\br,\br';\br_{N_e+1},\ldots,\br_{N_e+N_n}}}_n\right.\nonumber\\
&+\left.U_n\rbr{\br,\br'}\an{g_n\rbr{\br,\br';\br_{1},\ldots,\br_{N_e}}}_e}+\int_V d\mbf{r}'U_{e-n}\rbr{\br,\br'}g_{e-n}\rbr{\br,\br'} \nonumber\\
&+\rbr{v_{e}\rbr{\br}+c}\an{\rho_e\rbr{\br;\br_{N_e+1},\ldots,\br_{N_e+N_n}}}_n+\left.\rbr{v_{n}\rbr{\br}+c}\an{\rho_n\rbr{\br;\br_{1},\ldots,\br_{N_e}}}_e} d\mbf{r}=\abs{\rbr{E+c}N},
\end{align}
which is the analogue of Eq.~(\ref{pot_norm2}) for the Hamiltonian~(\ref{nuc_hamiltonian}) and the generalisation of $D_{v_2}$ to the external potential
$\sum_{i=1}^{N_e} v\rbr{\mbf{r}_{i}}+\sum_{i=1}^{N_n} v\rbr{\mbf{r}_{i}}$ is
\begin{equation} \label{pot_metric2_e_n}
 D_{v_{2},e-n}=\int\abs{h_{1,e-n}\ofr-h_{2,e-n}\ofr}d\mbf{r},
\end{equation}
where $h_{i,e-n}\rbr{\br}$ corresponds to the integrand of Eq.~(\ref{pot_norm2.2_e_n}) for system $i$. As was the case for $D_{v_{1}}$, the metric $D_{v_{2}}$ can be
generalised to measure the distance between systems containing an arbitrary number of sets of different particles $p_a,p_b,\ldots,p_m$ (e.g., systems which include electrons
and various ionic species). In this case, however, it is not required that corresponding ensembles of particles in different systems have the same size.
\twocolumngrid

\bibliography{References}
\end{document}